\documentclass[aps,pra,twocolumn,groupedaddress,superscriptaddress,showpacs]{revtex4}
\usepackage{natbib}
\usepackage[latin1]{inputenc}
\usepackage[dvips]{graphicx}
\usepackage{amsmath,amsbsy,amsfonts,amssymb}
\usepackage{bm}

\newcommand{\unit}[1]{\ensuremath{\,\rm #1}}
\newcommand{\rcm}{\mbox{cm$^{-1}$}}

\begin{document}
\title{The X$^1\Sigma^+$ and a$^3\Sigma^+$ states of LiCs studied by
Fourier-transform spectroscopy}
\author{Peter Staanum}
\altaffiliation{Now at Institute of Physics and Astronomy,
University of Aarhus, Denmark.} \affiliation{Institut für
Quantenoptik, Gottfried Wilhelm Leibniz Universität Hannover,
Welfengarten 1, 30167 Hannover, Germany}
\author{Asen Pashov}
\affiliation{Department of Physics, Sofia University, 5 James
Bourchier Boulevard, 1164 Sofia, Bulgaria}
\author{Horst Knöckel}
\affiliation{Institut für Quantenoptik, Gottfried Wilhelm Leibniz Universität Hannover,
Welfengarten 1, 30167 Hannover, Germany}
\author{Eberhard Tiemann}
\affiliation{Institut für Quantenoptik, Gottfried Wilhelm Leibniz Universität Hannover,
Welfengarten 1, 30167 Hannover, Germany}
\received{\today}
\begin{abstract}
We present the first high-resolution spectroscopic study of LiCs.
LiCs is formed in a heat pipe oven and studied via laser-induced
fluorescence Fourier-transform spectroscopy. By exciting molecules
through the X$^1\Sigma^+$-B$^1\Pi$ and X$^1\Sigma^+$-D$^1\Pi$
transitions vibrational levels of the X$^1\Sigma^+$ ground
state have been observed up to 3 \unit{cm^{-1}} below the dissociation limit enabling an accurate construction
of the potential. Furthermore, rovibrational levels in the
a$^3\Sigma^+$ triplet ground state have been observed because the excited states obtain sufficient triplet character at the corresponding excited atomic asymptote. With the help of coupled channels calculations accurate singlet and triplet ground state potentials were derived reaching the atomic ground state asymptote and allowing first predictions of cold collision properties of Li + Cs pairs.

\end{abstract}
 \maketitle
\section{Introduction}

Spectroscopy of heteronuclear diatomic alkali molecules provides
important input to current research in cold molecules and mixtures
of ultracold atomic gases. Cold heteronuclear alkali dimers are
subject to a large interest since they can be formed at
temperatures below
1\unit{mK}~\cite{Kraft_LiCs2006,Mancini-KRbmolecules,Wang-KRb,Haimberger-groundstate,Kerman-RbCs}
and possess a large permanent electric dipole
moment for deeply bound singlet levels~\cite{Aymar2005,Igel-Mann_1986}. This combination of
properties enables electric field control of ultracold
collisions~\cite{Krems2005,Krems2006} and cold chemical
reactions~\cite{Balakrishnan2001,Bodo2002} and holds promises for
applications in quantum computation~\cite{DeMille2002}. Precise
potential energy curves, in particular for the lowest electronic
states, are evidently important for such applications as well as
for understanding the molecule formation processes
(photoassociation), ro-vibrational state selective
detection~\cite{Kerman-RbCs,Wang-Rempi2005} and for formation of
vibrational ground-state molecules~\cite{Sage-RbCs-v0}. In
ultracold mixtures of atomic gases, quantum degeneracy of one
atomic species can be achieved through sympathetic cooling by the
other species~\cite{DeMarco_FermiDeg1999,Modugno_K-RbBEC2001,Truscott_Li6Li7_2001}.
The interspecies interaction strength can be varied through
magnetic Feshbach resonances and has a large influence on a
variety of effects such as phase-separation between a
Bose-Einstein condensate and a degenerate Fermi gas
~\cite{Molmer_PhaseSep_PRL1998} and the transition to a
Bardeen-Cooper-Schrieffer superfluid state in dilute Fermi
gases~\cite{Heiselberg_Superfluid_PRL2000}. Understanding
interspecies collision properties at the atomic ground state
asymptotes, e.g. predicting the magnetic field strength values of Feshbach
resonances at zero kinetic energy, requires precise potential
curves of the electronic ground states.

Cold LiCs molecules were observed very recently, being formed in a
two-species magneto-optical trap~\cite{Kraft_LiCs2006}.
Previously, inelastic collisions~\cite{Schloeder} and sympathetic
cooling of Li by Cs in an optical dipole
trap~\cite{mudrich2002:prl} have been studied. Recent theoretical
work considers LiCs in strong dc electric fields and its influence
on rovibrational dynamics~\cite{Gonzalez-Ferez2006} and Li-Cs
collision cross sections~\cite{Krems2006}.

\begin{figure}
  \centering
  \includegraphics[width=\linewidth]{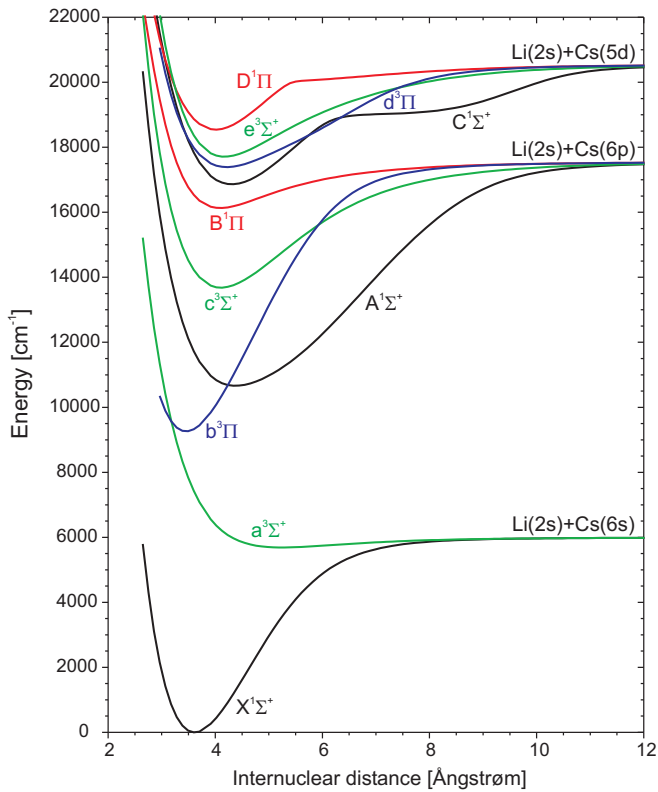}
  \caption{(Color online) Selected singlet and triplet potential energy curves in LiCs~\cite{Korek_2000}.}\label{fig:potentials}
\end{figure}

Although LiCs was observed already in 1928 by Walter and Barratt
through absorption in a mixture of metallic
vapors~\cite{Walter_1928}, very few and no high-resolution
spectroscopic studies have been made until now. In the 1980's
Vadla \emph{et al.} studied the repulsive Li(2p)+Cs(6s)
asymptote~\cite{Vadla_1983}. More recently LiCs molecules were
formed on He nanodroplets and excitation spectra of the
d$^3\Pi\leftarrow$ a$^3\Sigma^+$ transition were recorded and
modeled~\cite{Mudrich_2004}. Ab initio potentials were calculated
by Korek \emph{et al.}~\cite{Korek_2000} (see
Fig.~\ref{fig:potentials}) and more recently an extended
theoretical study was done by Aymar and Dulieu~\cite{Aymar2005}.
The theoretical potentials provide a good starting point for
analysis of the spectra obtained in the present work.

Here we present a high-resolution spectroscopic study of the LiCs
molecule. Similar to our previous studies
\cite{pashov:05,Docenko_NaCs2006} we apply Fourier-transform
spectroscopy of laser-induced fluorescence from LiCs molecules
formed in a heat-pipe, because this technique is suitable for
collecting a large amount of accurate experimental data. By using
properly chosen excitation schemes (see e.g.
\cite{pashov:05,Docenko_NaCs2006}) we can measure transition
frequencies to a wide range of vibrational levels in the singlet
X$^1\Sigma^+$ as well as in the triplet a$^3\Sigma^+$ ground states,
especially levels close to the asymptote. Since the two states are
coupled at long internuclear distances by the hyperfine interaction
it is not correct to treat them separately in this region, as it
would lead to model potentials which are unable to reproduce the
experimental observations close to the asymptote. Therefore, the aim
of our experimental work is to collect experimental data on both
ground states and to fit accurate experimental potential energy
curves simultaneously for both states - the indispensable starting
point for a study of the molecular structure of LiCs or for modeling
of cold collision processes on the Li(2s)+Cs(6s) asymptote. We apply
these potential curves within a coupled channels model in order to
explain our experimental observations and also to compute collision
properties for comparing recent results of sympathetic cooling of Li
by Cs \cite{mudrich2002:prl}

The article is organized as follows: The experimental setup and
excitation schemes are presented in Section~\ref{sec:Experimental}.
In Section~\ref{sec:Data Analysis} we describe the analysis of the
obtained spectra. The procedure for construction of potential energy
curves is described in Section~\ref{sec:Potential construction} and
the potentials are reported. In Section~\ref{sec:conclusion} we give
our conclusion and an outlook for further experimental study needed for a quantitative description of ultracold collisions in Li + Cs.

\section{Experiment: molecule formation and spectroscopy}\label{sec:Experimental}
\subsection{Molecule formation}
LiCs molecules are formed in a stainless steel heat-pipe identical
to the one described in Ref.~\cite{Docenko_NaCs2004}, except for one
modification described below. The heat-pipe (960\unit{mm} long and
34\unit{mm} outer diameter) is filled with 6\unit{g} Li and
5\unit{g} Cs (the Cs, in a closed ampoule, is loaded into the side
container~\cite{Docenko_NaCs2004}~\footnote{Note that the CF flange
on the side container~\cite{Docenko_NaCs2004} is here closed using a
nickel gasket; traditional copper gaskets are corroded by
Li~\cite{Stan-DoubleOven}.}) and typically operated with 3-6
\unit{mbar} Ar buffer gas pressure.

Since the vapor pressure difference between Li and Cs at a common
temperature is extremely large, we modified the design of
Ref.~\cite{Docenko_NaCs2004} in order to obtain a three-section
heat-pipe, which is more suitable for producing a vapor mixture
with similar concentrations of Li and Cs and hence for forming
LiCs molecules~\cite{Bednarska_MeasSciTech1996}. In
Ref.~\cite{Docenko_NaCs2004} the central 60\unit{cm} of the
heat-pipe is heated uniformly in a commercial oven (Carbolite);
here we mount two stainless steel 'shells' (20\unit{cm} long,
50\unit{mm} inner diameter) concentrically around the heat-pipe
and seal up the ends facing towards the center of the oven such
that only the central 20\unit{cm} of the heat-pipe are heated
directly in the oven. By blowing air into the open ends, which
extend outside of the oven, we can maintain a lower temperature in
the sections shielded by the shells than in the central part of
the heat-pipe.

The heat-pipe is conditioned by heating it to temperatures of
about 580\unit{^\circ C} under 10\unit{mbar} Ar pressure;
subsequently the Cs ampoule is broken by shaking the tube.
Operating temperatures are 540\unit{^\circ C} in the central part
and 370\unit{^\circ C} in the outer sections. The heat-pipe oven
was operated for more than 200 hours over a 10 month period and
was still in good working conditions at the end of this period.

\subsection{Laser-induced fluorescence Fourier-transform spectroscopy}

Laser-induced fluorescence from LiCs molecules is observed after
excitation on the B$^1\Pi\leftarrow$ X$^1\Sigma^+$ and
D$^1\Pi\leftarrow$ X$^1\Sigma^+$ transitions. The
B$^1\Pi\leftarrow$ X$^1\Sigma^+$ transitions were excited using a
Coherent 599 dye laser (with DCM dye) at frequencies in the range
15529-16123\unit{cm^{-1}} and a Coherent 699 dye laser at
frequencies in the range 16397-17022\unit{cm^{-1}} (with Rhodamine
6G dye as well as with a mixture of Rhodamine 6G and Rhodamine B).
The D$^1\Pi\leftarrow$ X$^1\Sigma^+$ transition was excited using
the dye laser with Rhodamine 6G at frequencies in the range
16663-17238\unit{cm^{-1}} and a frequency doubled Nd:YAG laser.
The strongest signals were observed for the B$^1\Pi\leftarrow$
X$^1\Sigma^+$ system and we studied the ground states mainly
through this system. We note that indeed Walter and Barratt
observed strong absorption in the range
15983-16582\unit{cm^{-1}}~\cite{Walter_1928} which according to
Fig.~\ref{fig:potentials} corresponds to the B$^1\Pi\leftarrow$
X$^1\Sigma^+$ transition. At excitation frequencies in the range
$\sim 14900-15400$\unit{cm^{-1}}, we find strong fluorescence due
to Li$_2$ and NaCs (Na is present as an impurity in the Li sample)
which overshadows a possible LiCs signal. Using an Ar-ion laser
for excitation at 457.9\unit{nm}, 476.5\unit{nm}, 488.0\unit{nm},
496.5\unit{nm} and 514.5\unit{nm} we did not observe any
fluorescence from LiCs molecules, only from Li$_2$, LiNa and NaCs.

Contrary to the previously studied molecules NaRb~\cite{pashov:05}
and NaCs~\cite{Docenko_NaCs2006}, the low lying levels of the
excited B$^1\Pi$ state in LiCs turned out to be almost free of
local perturbations by the neighboring triplet states b$^3\Pi$ and
c$^3 \Sigma^+ $, which could be expected from the theoretical
potential curves (see Fig.~\ref{fig:potentials}). As a consequence
we were unable to register any transition to the triplet ground
state from the low lying B$^1\Pi$ state levels. We searched
instead for access to the triplet manifold through local
perturbations in the D$^1\Pi$ and C$^1\Sigma^+$ states. Here we
used a Coumarine 6 dye laser with a typical power of 25 mW.
Unfortunately within the searched excitation frequency region of
18446 - 19039 \rcm\ we were also not able to register transitions
to the triplet ground state.

The only fluorescence to the a$^3\Sigma^+$ state is observed after
excitation to high lying levels in the B$^1\Pi$ state. These
transitions are attributed to the long-range change over of coupling
case for the B$^1\Pi$ state itself rather than to local mixing of
the B$^1\Pi$ state at long range with the neighboring triplet
states~\cite{pashov:05,Docenko_NaCs2006}. This conclusion is
supported first, by the observation that high lying B state levels
seem to be locally unperturbed. Second, the intensity distribution
of progressions from the high lying B state levels to the ground
singlet state could be explained satisfactory by the Franck-Condon
factors between the X and the B states, including the highest
$v_{X}''$ levels (contrary to the case of NaRb \cite{pashov:05}). So
transitions from these B state levels to high lying a$^3\Sigma^+$
state levels will become also probable when the B state changes its
character from Hund's case (a) $^1\Pi$ to Hund's case (c) $\Omega=1$
at long internuclear distances. Indeed, in our spectra we find
transitions mainly to high $v_{a}''$ and none to the bottom of the
triplet ground state.

The laser-induced fluorescence light is collected in the direction
opposite to the one of laser beam propagation and recorded by a
Bruker IFS 120HR Fourier-Transform Spectrometer (FTS). For detection
we use a photomultiplier (Hamamatsu R928) or a Si-photodiode. In
order to avoid illumination of the detector by the He-Ne laser
(632.8 nm), used in the FTS for calibration and stepping control, a
notch filter (8\unit{nm} full width at half maximum) is introduced
in the beam path, which suppresses also the fluorescence induced in
the corresponding spectral region. The resolution of the FTS is
typically set to 0.03 - 0.05 \unit{cm^{-1}}. The uncertainty of the
line positions is estimated to be $1/10$ of the resolution. For
lines with signal-to-noise ratio less than 3 the uncertainty is
gradually increased. Each spectrum results typically from an average
of 10-20 scans, but the number of scans is varied from 5 to 350
depending on the signal strength for the features of interest in the
spectra. For improving the signal-to-noise ratio, the spectral
window for some spectra is limited by using color glass filters or
interference filters. In order to facilitate the identification in
such cases we recorded also a spectrum at the same excitation
frequency without filters, but with lower resolution (0.1 \rcm) and
a smaller number of scans.

A list of all excitation frequencies used in these experiments
together with the assignment of the excitation transitions are given
in Tables 3 and 4 of the supplementary materials~\cite{EPAPS}.

\section{Analysis of spectra}\label{sec:Data Analysis}
\subsection{The X$^1\Sigma^+$ state}

Assignment of the large number of transitions, about 6600, to
the X$^1\Sigma^+$ state is done in an iterative process; by
gradually improving the potential for the X$^1\Sigma^+$ state,
more and more transitions can be correctly assigned.

Initially, we identify several strong doublet series which are
clearly recognized as P-R components from the regular development
of the doublet spacing. Among these series we choose those with
similar spacing, i.e., with similar rotational quantum numbers.
Based on the theoretical potential for the X$^1\Sigma^+$
state~\cite{Korek_2000} we make an initial guess for the
vibrational and the rotational quantum numbers of these
fluorescence progressions. The quantum numbers that give the
closest agreement with the theoretical vibrational and rotational
spacings are assigned and a small set of Dunham coefficients is
fitted. If the fit is successful we can use the fitted
coefficients to assign new progressions; if not, a reassignment of
the quantum numbers must be made. After some iterations we obtain
a self consistent set of assigned experimental progressions which
can be satisfactory described by a few Dunham coefficients. This
is a first hint of a correct rotational numbering. Here we point
out the very good quality of the theoretical calculations
\cite{Korek_2000,Aymar2005}; the theoretical rotational numbering
needed a correction by only two or three units.

When the list of assigned transitions reaches several hundreds we
perform the first potential fits. Initially, we start with a
pointwise potential (defined in Section~\ref{sec:Potential
construction}) based on the theoretical curve \cite{Korek_2000}
and improve it using the procedure described in
Ref.~\cite{ipaasen}. In the further analysis of the spectra we
apply this pointwise potential curve since outside the range of
the fitted $v''$ and $J''$, it usually possesses better predictive
properties than the Dunham type coefficients. Moreover our
experience from previous studies \cite{Docenko_NaCs2004,pashov:05}
shows that a potential curve which fits long vibrational
progressions recorded at high precision in a wide range of
rotational quantum numbers indicates the correctness also of the
vibrational numbering.

Finally, the established vibrational numbering was confirmed by
assigning several progressions for the less abundant
$^6$Li$^{133}$Cs molecule. These progressions fit to the
experimental potential based only on $^7$Li$^{133}$Cs data when
the appropriate reduced mass is applied \cite{mass2003}.

In order to describe the important long-range part of the potential,
it is of great value to collect data with transitions to high-lying
levels of the ground state. Since we noticed during the measurements
that transitions to such high-lying levels originate from high-lying
vibrational levels of the B$^1\Pi$ state, we studied this state in
order to optimize the experimental conditions for observation of
near asymptotic levels in the X$^1\Sigma^+$ state (a report on the
B$^1\Pi$ state is in preparation~\cite{pashovBstate}). From the
available experimental term energies of the B$^1\Pi$ state levels a
preliminary potential for the B$^1\Pi$ state was fitted. This
potential was then used to predict transition frequencies for
excitation transitions with large Franck-Condon factors. In this way
we recorded systematically transitions from $v'=24$ and 25 to the
X$^1\Sigma^+$ state for a wide range of $J'$, thus adding several
ro-vibrational levels with $v_{X}''=49$ and 50 to the ground state
dataset. In most cases such high-lying levels of the X$^1\Sigma^+$
state could be observed with a sufficient signal-to-noise ratio only
by increasing the number of summed scans to several hundreds. We
searched for excitations to higher vibrational levels of the
B$^1\Pi$ state within predicted spectral regions but we were not
able to register fluorescence from $v' \geq 26$ to the X$^1\Sigma^+$
state, most likely due to possible predissociation or unfavorable
transition probabilities.

\subsection{The a$^3\Sigma^+$ state}

\begin{figure} 
  \centering
  \includegraphics[width=\linewidth]{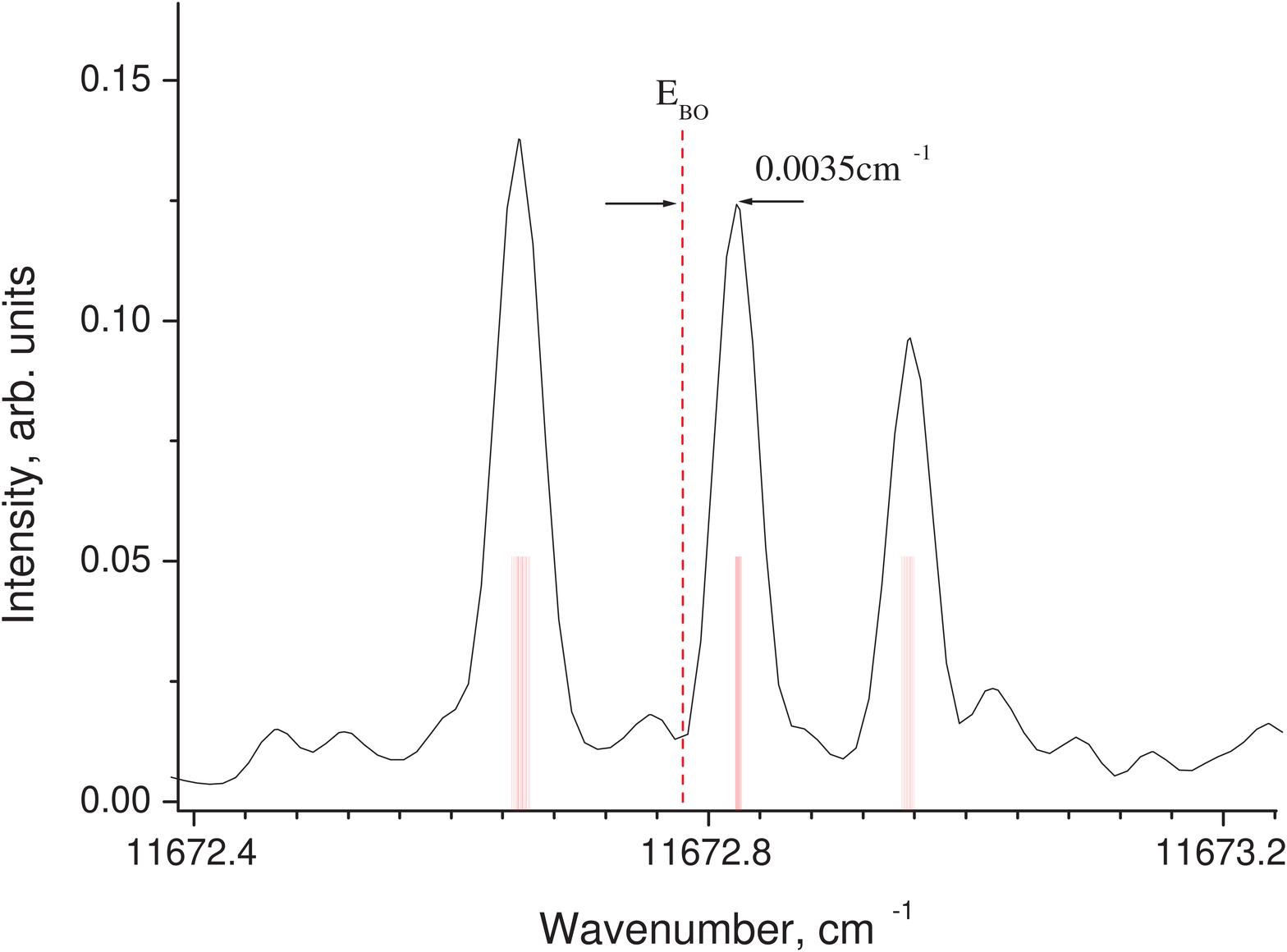}
  \caption{(Color online) Hyperfine structure of a transition to the
  $v_a''=9$, $N_{a}''=17$ level of the a$^3\Sigma^+$ state.
  The vertical full lines indicate the prediction
  of the coupled channels calculation. The dashed vertical line
  indicates the position of the hyperfine-structure free level, which is
  shifted by 0.035 \rcm\ from the central component of the structure.}
\label{HFS}
\end{figure}
\begin{figure*}[htp]
  \centering
  \includegraphics[width=\linewidth]{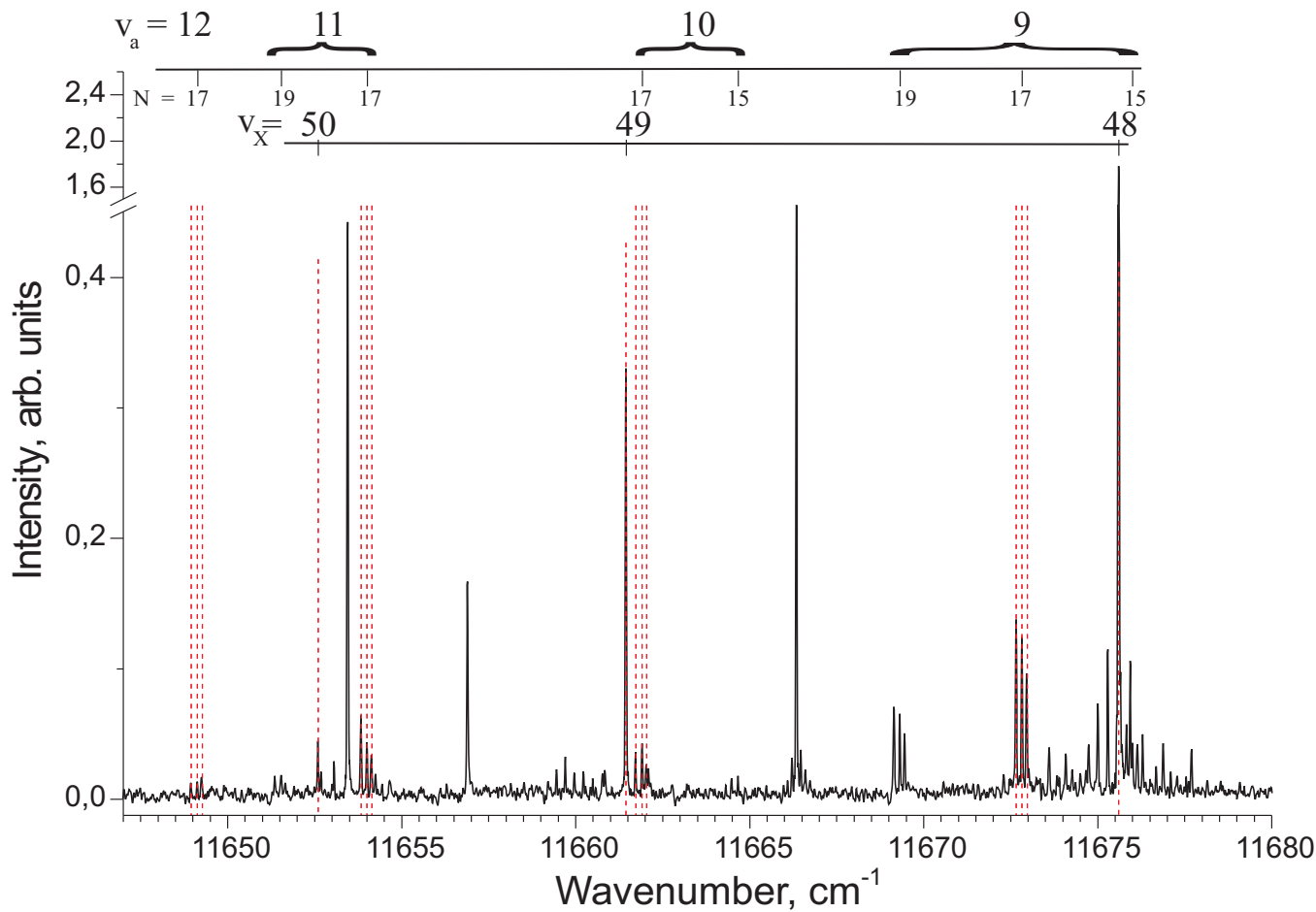}
  \caption{(Color online) Portion of recorded progression from the B state level
with $v'=25$, $J'=17$ excited by a Q-type transition. Lines not
marked on the figure belong to another assigned progression.}
\label{j17}
\end{figure*}

The triplet transitions are easily distinguished by their hyperfine
structure (HFS) which at our resolution consists of three lines
split by approximately 0.3 \rcm. Calculations similar to those in
Ref.~\cite{Kasahara:96} show that the observed splitting is well
reproduced by the Fermi contact interaction model applying the
atomic HFS constants for $^7$Li and $^{133}$Cs \cite{Arimondo}. In
Fig.~\ref{HFS} a transition to the triplet state level $v''_{a}=9,
N_{a}''=17$ \footnote{Here $N$ is the rotational quantum number for a
Hund's case (b) state and the total angular momentum $\bm{J}$ is
the sum of $\bm{N}$ and the total electron spin $\bm{S}$.} is shown
together with the prediction of the splitting modeled by a coupled
channels calculation as described in Section~\ref{sec:Potential
construction}. A larger portion of the same progression originating
from the $v'=25$, $J'=17$ B$^1\Pi$ state level excited by a Q-type
($\Delta J=0$) transition is shown in Fig.~\ref{j17}. Some of the
transitions reach near asymptotic levels in both ground states. The
vertical dashed lines indicate the prediction of the coupled
channels calculation for $J''=N''=17$ in which the hyperfine
interaction between the states is included. The progression to the
X$^1\Sigma^+$ state is formed by Q-lines whereas the progression to
the a$^3\Sigma^+$ state consists of transitions to $N_a''=$15, 17
and 19. Lines not marked in the figure belong to another progression
also assigned and used in our analysis.

The rotational assignment of the triplet lines is straightforward
since we always find the progression to the X$^1\Sigma^+$ state
which shares the excited level in the B$^1\Pi$ state with the
progression to the a$^3\Sigma^+$ state. The vibrational assignment
of the transitions to the triplet ground state is done in the same
way as for the X$^1\Sigma^+$ state, however, no transitions in
$^{6}$Li$^{133}$Cs were observed and hence the vibrational
assignment relies only on the internal consistency of the total
procedure.

\subsection{Data sets}
\begin{figure*}
  \centering
  \includegraphics[width=0.48\linewidth]{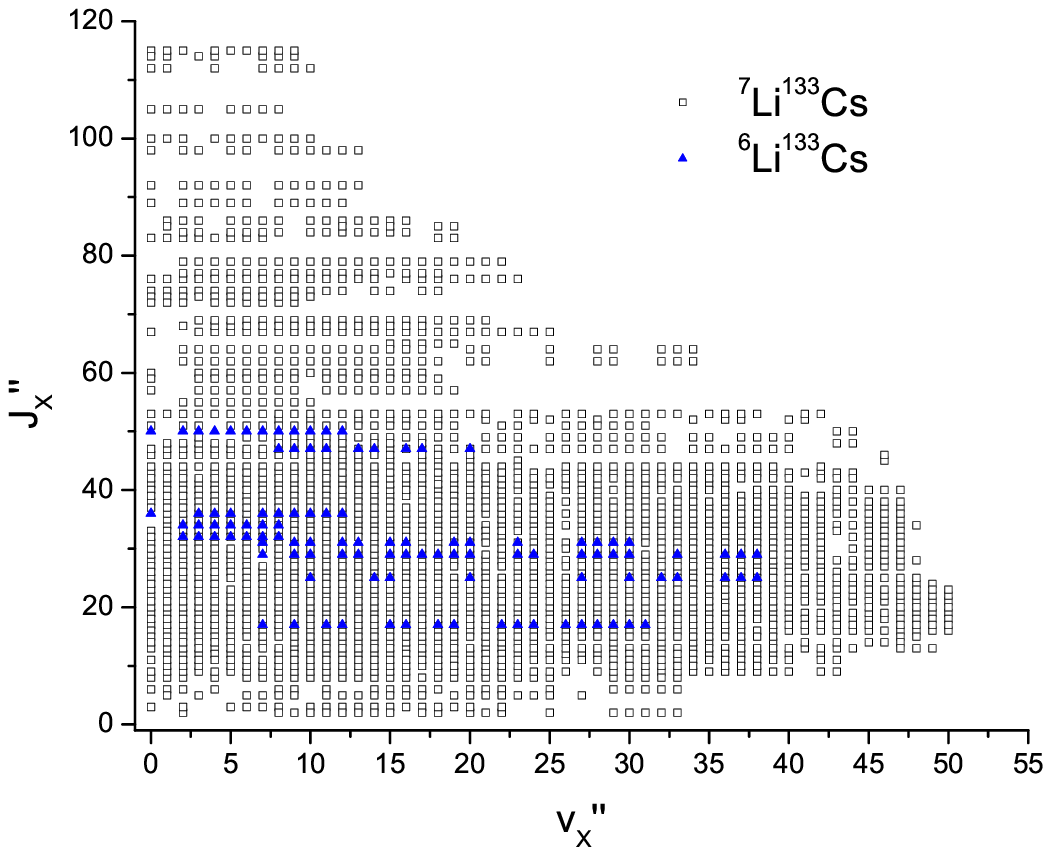}
  \includegraphics[width=0.48\linewidth]{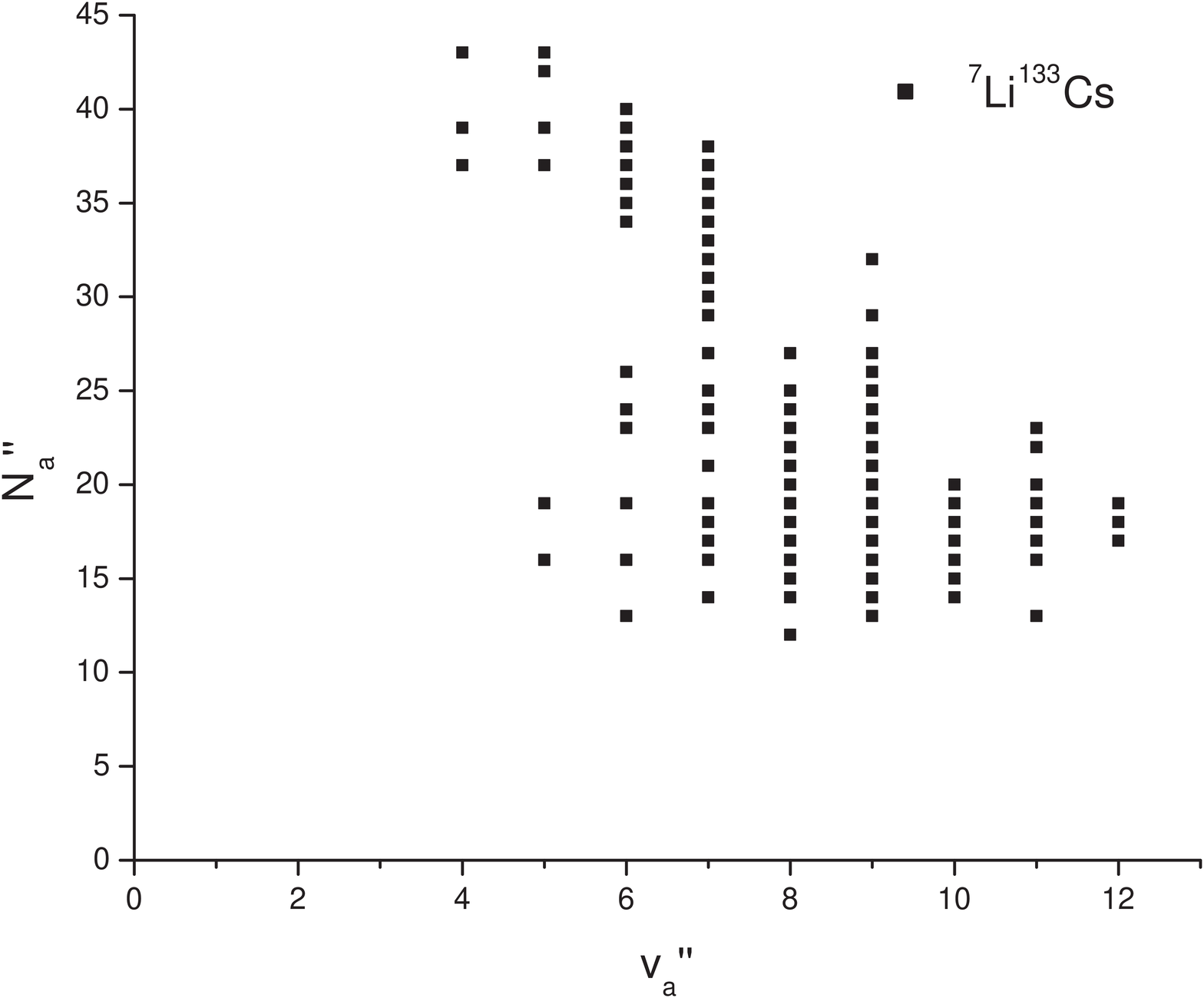}
  \caption{(Color online) Observed ro-vibrational levels in the X$^1\Sigma^+$ and a$^3\Sigma^+$ states of $^7$Li$^{133}$Cs and $^6$Li$^{133}$Cs.} \label{datafields}
\end{figure*}
Altogether about 6600 transitions to the X$^1\Sigma^+$ state (130 of
them belong to $^6$Li$^{133}$Cs) and 180 transitions to the
a$^3\Sigma^+$ state in $^7$Li$^{133}$Cs were assigned (see Tables 1
and 2 of the supplementary materials~\cite{EPAPS}). The
corresponding distribution of vibrational and rotational quantum
numbers is shown in Figure~\ref{datafields}. For the X$^1\Sigma^+$
state we observe transitions to $v_X''=0-50$ in $^7$Li$^{133}$Cs.
The set of a$^3\Sigma^+$ state vibrational levels covers
$v_a''=4-12$ and the vibrational numbering established in this study
agrees with that from the theoretical potential. The limited number
of experimental data, however, might lead to a revision of this
assignment in the future, if additional data on the a$^3\Sigma^+$
state are collected. Although the data set for the a$^3\Sigma^+$
state may seem fragmentary compared to our similar studies in other
molecules, the collection of these data is extremely valuable,
first, due to the limited possibilities for exciting triplet states
in LiCs and, second, since a proper description of the Li(2s)+Cs(6s)
asymptote is only possible if both the X$^1\Sigma^+$ state and
a$^3\Sigma^+$ state are treated in a coupled channels manner as
described in the next section.

\section{Construction of potential energy curves}\label{sec:Potential construction}

The self-consistent assignment and fitting procedure described
above gives rise to accurate pointwise short-range
potentials~\cite{ipaasen}. For the long-range part of both
potentials we use an extension of the form
\begin{equation}
U_{\mathrm{LR}}(R) =U_{\infty} - \frac{C_{6}}{R^{6}}-
\frac{C_{8}}{R^{8}} - \frac{C_{10}}{R^{10}} \pm E_{ex} \mbox{ .}
\label{LRs}
\end{equation}
Here $U_{\infty}$ is the energy of the atomic asymptote with respect
to the minimum of the X$^1\Sigma^+$ state potential, $C_6$, $C_8$
and $C_{10}$ are the dispersion coefficients and
\begin{equation}
E_{ex}(R) = A_{ex} \cdot R^\gamma \cdot e^{-\beta R}
\label{exchange}
\end{equation}
is the frequently applied functional form of the exchange energy
\cite{Smirnov} which is added for the triplet state and subtracted
for the singlet state.

The pointwise short-range and the long-range potentials are
connected at a point $R_{o}$ ensuring a smooth transition between
both potential branches. $R_o$ is chosen as described in
Refs.~\cite{Allard_Ca2_2002,pashov:05}, the $C_6$ and $C_8$
coefficients are fixed to their theoretical values
\cite{Marinescu_1999, Derevianko_C62001, Porsev:03}, $\gamma$ and
$\beta$ are estimated using the ionization potentials for Li and
Cs \cite{Radzig} according to Ref.~\cite{Smirnov}, while
$U_{\infty}$, $C_{10}$ and A$_{ex}$ are adjusted during the
fitting procedure.

In the first step of the fitting procedure we adjust only the
pointwise part of the potential for the
X$^1\Sigma^+$ state. The experimental transition frequencies are fitted
by adjusting the parameters of the pointwise potential
and the term energies of the excited levels.

As we choose the origin of the potential energy at the minimum of
the X$^1\Sigma^+$ state potential, in the second step of the fitting
procedure we need to determine the term energies of the
a$^3\Sigma^+$ state levels with respect to the X$^1\Sigma^+$ state.
We do this using spectra where we observe simultaneously
progressions to the singlet and the triplet states originating from
a common upper state level. For a given singlet ground state
potential we calculate the term energies of the triplet state using
the energy of the excited level, determined in the previous step,
and the progression to the a$^3\Sigma^+$ state from this level.
These a$^3\Sigma^+$ term energies are then used in order to fit the
potential parameters of the triplet state. In this way we ensure
always a proper position of the a$^3\Sigma^+$ state with respect to
the X$^1\Sigma^+$ state.

In order to treat the hyperfine structure of the spectral lines of the triplet state we checked the experimental data that the splitting within our resolution is independent of the vibrational
and rotational quantum numbers $v''$ and $N''$ (except for very few
cases, which we discuss below). Therefore, we use the central
component of the structure for identification of the transition (see
Fig.~\ref{HFS}). We convert the observed frequency to term energy, and
take into account the shift (-0.035 \rcm, which means the hyperfine level is more deeply bound than the unperturbed level) of the selected hyperfine component from the unperturbed, hyperfine structure free one, thus we fit with the constructed term values a Born-Oppenheimer potential.

\begin{figure*}[htp]
  \centering
  \includegraphics[width=\linewidth]{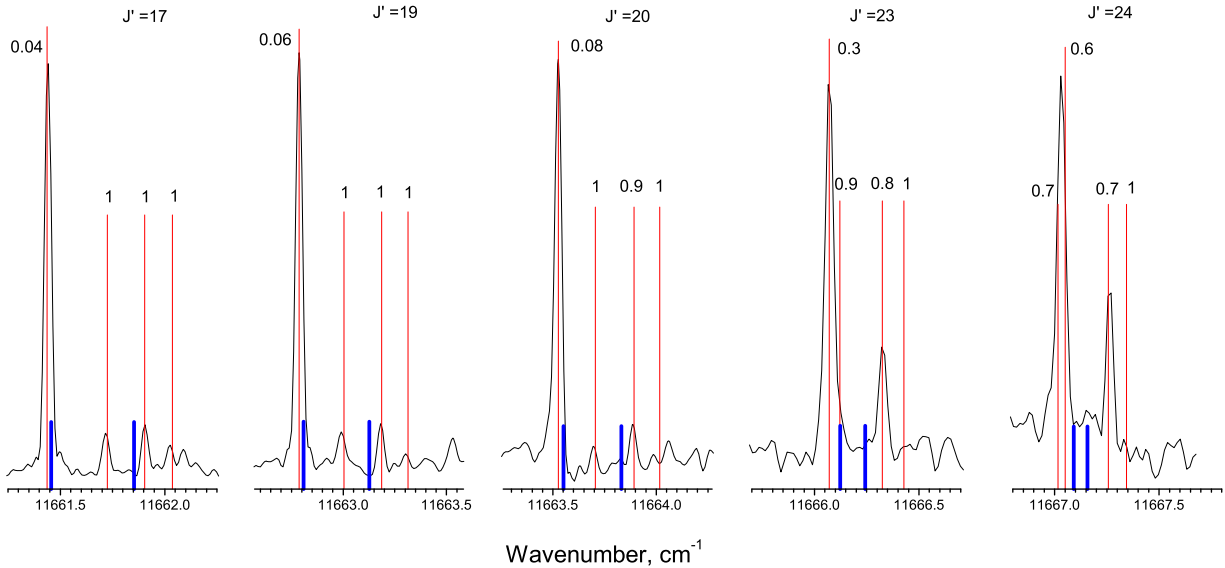}
  \caption{(Color online) Observed transitions to $v_X''=49$ and $v_a''=10$
   for different $J'$ as indicated on the figure and excited by a Q-type transition.
   An extended portion of the spectrum for $J'=17 $ for several $v_a''$ is shown in Fig.~\ref{j17}.
   For increasing $J'$ the singlet  and triplet lines approach each
   other, resulting in strong perturbations of the line structure and the intensity
   distribution within the HFS of the triplet line. Thick blue bars indicate the line
   positions as predicted by the Born-Oppenheimer potentials. Thin red bars indicate
   the prediction of the coupled channels calculation. The
   numbers close to these bars are the corresponding expectation values of the total spin operator $\left|\bm{S}\right|$.
   For convenience spectra are adjusted such that the largest peaks have similar intensities.}
\label{hfs_comp}
\end{figure*}

For high vibrational levels, and especially in cases of close
approach of singlet and triplet levels with the same rotational
quantum numbers, significant deviations from the Born-Oppenheimer
picture can be expected \cite{pashov:05,Docenko_NaCs2006}. In
$^7$Li$^{133}$Cs this is most pronounced for $v_{X}''=49$ in the
singlet and $v_{a}''=10$ in the triplet state (see Fig.~\ref{j17}).
The deviations of the transitions determined by the Born-Oppenheimer
potentials (indicated with thick blue bars in Fig.~\ref{hfs_comp})
from the experimental ones are significant and reach 0.06 \rcm\ for
$v_{X}''=49$ and $J'=24$. Fig.~\ref{hfs_comp} shows the development
from low $J$ to high $J$.

Therefore, in a third step, the Born-Oppenheimer potentials are
refined by extending the single channel approach from the first two
steps and applying a coupled channels calculation as discussed in
detail in Refs.~\cite{pashov:05,Docenko_NaCs2006}. Briefly, we
calculate the difference between the single channel and coupled
channels eigenvalues and subtract these differences from the
experimentally observed transition frequencies in order to obtain
the frequencies which would be observed without hyperfine coupling
between the singlet and the triplet ground states. Next, these
frequencies are used in a combined fit to adjust the parameters of
the long-range extensions of the potentials as well as in separate
fits (described for the first two steps) of the short range
pointwise part of the triplet and singlet potentials. Finally, the
whole fitting procedure is repeated until the frequencies predicted
with the coupled channels calculations agree with the experimental
observations. The importance of the coupled channels calculation is
illustrated in Fig.~\ref{hfs_comp}, where the predictions for the
coupled system are shown with thin red bars together with the
experimentally observed lines and the single channel predictions
(thick blue bars). The degree of mixing between the triplet and
singlet levels is characterized by the expectation value of the
total spin operator $\left|\bm{S}\right|$ which is indicated for
each line in Fig.~\ref{hfs_comp}.

\begin{table}[htp]

  \centering
  \caption{Pointwise representation of the potential energy curve for
 the X$^1\Sigma^+$ state of LiCs. See also Table 5 of the supplementary materials~\cite{EPAPS}.}
 \begin{tabular*}{0.9\linewidth}{@{\extracolsep{\fill}}rrrr}\hline
 R [\AA] & U [cm$^{-1}$]& R [\AA] & U [cm$^{-1}$]\\ \hline
  2.110000 &       23121.37091 &  4.898338 &        2503.26403 \\
  2.193910 &       18288.09701 &  5.035379 &        2854.02395 \\
  2.277810 &       14589.99494 &  5.172421 &        3188.23112 \\
  2.361720 &       11737.62964 &  5.309462 &        3502.00098 \\
  2.445630 &        9497.25485 &  5.446503 &        3792.60234 \\
  2.529540 &        7805.39559 &  5.583500 &        4058.26131 \\
  2.613440 &        6434.86554 &  5.763111 &        4367.79625 \\
  2.697350 &        5258.36144 &  5.942722 &        4633.88339 \\
  2.781259 &        4228.82452 &  6.122333 &        4858.92343 \\
  2.865166 &        3334.95455 &  6.301944 &        5046.60206 \\
  2.949073 &        2568.92046 &  6.481556 &        5201.36003 \\
  3.032980 &        1922.48746 &  6.661167 &        5327.82056 \\
  3.116800 &        1387.62221 &  6.840778 &        5430.48784 \\
  3.253841 &         728.62698 &  7.020389 &        5513.45725 \\
  3.390883 &         302.73498 &  7.200000 &        5580.28406 \\
  3.527924 &          71.68736 &  7.550000 &        5675.34592 \\
  3.664966 &            .03204 &  7.900000 &        5737.67458 \\
  3.802007 &          56.01497 &  8.250000 &        5778.85417 \\
  3.939048 &         211.90321 &  8.600000 &        5806.38135 \\
  4.076090 &         443.87130 &  8.950000 &        5825.08952 \\
  4.213131 &         731.70689 &  9.300000 &        5837.98033 \\
  4.350172 &        1058.39492 & 10.225000 &        5856.82445 \\
  4.487214 &        1409.72907 & 11.150000 &        5865.32302 \\
  4.624255 &        1773.86942 & 12.075000 &        5869.51458 \\
  4.761297 &        2141.05773 & 13.000000 &        5871.76955 \\
&&&\\
\multicolumn{2}{l}{$U_{\infty}$=5875.45504 \rcm} & & \\
\multicolumn{2}{l}{$R_{\mathrm{o}}$=11.5275 \AA} & & \\
\multicolumn{2}{l}{$C_6$=1.47714$\cdot 10^{7}$ \rcm\AA $^{6}$ } &
\multicolumn{2}{l}{$A_{ex}$=3.81419$\cdot 10^{4}$ \rcm\AA $^{-\gamma}$} \\
\multicolumn{2}{l}{$C_8$=4.33209$\cdot 10^{8}$ \rcm\AA $^{8}$} &
\multicolumn{2}{l}{$\gamma$=5.0568} \\
\multicolumn{2}{l}{$C_{10}$=1.21271$ \cdot 10^{10}$ \rcm\AA
$^{10}$}&
\multicolumn{2}{l}{$\beta$=2.2006 \AA $^{-1}$} \\
&&&\\
 \multicolumn{2}{l}{$T^{\mathrm X}_{\mathrm e}$=0 \rcm} &
{$R^{\mathrm X}_{\mathrm e}$=3.6681 \AA} &\\
 \multicolumn{2}{l}{$D^{\mathrm X}_{\mathrm e}$=5875.455(100) \rcm} &
 \multicolumn{2}{l}{$D^{\mathrm X}_{\mathrm 0}$=5783.408(100) \rcm} \\

 \\\hline
\end{tabular*}
\label{ipapotX}
\end{table}

\begin{table}[htp]

  \centering
  \caption{Pointwise representation of the potential energy curve for
 the a$^3\Sigma^+$ state of LiCs. See also Table 5 of the supplementary materials~\cite{EPAPS}.}
 \begin{tabular*}{0.9\linewidth}{@{\extracolsep{\fill}}rrrr}\hline
 R [\AA] & U [cm$^{-1}$]& R [\AA] & U [cm$^{-1}$]\\ \hline
       3.020000 &   10529.73474 &       7.673846 &    5783.12762 \\
       3.384521 &    8133.49216 &       8.093333 &    5805.67410 \\
       3.749042 &    6806.18429 &       8.512821 &    5822.85769 \\
       4.113562 &    6133.96772 &       8.932308 &    5835.76854 \\
       4.478083 &    5768.93998 &       9.597282 &    5849.85069 \\
       4.769700 &    5630.11787 &      10.190769 &    5857.88706 \\
       5.318264 &    5567.01710 &      11.000000 &    5864.74252 \\
       5.866825 &    5613.25953 &      12.000000 &    5869.36430 \\
       6.415385 &    5676.25455 &      13.000000 &    5871.79045 \\
       6.834872 &    5718.42584 &      14.000000 &    5873.16557 \\
       7.254359 &    5754.16017 & &\\
&&&\\
\multicolumn{2}{l}{$U_{\infty}$=5875.45504 \rcm} & & \\
\multicolumn{2}{l}{$R_{\mathrm o}$=11.5183 \AA} & & \\
\multicolumn{2}{l}{$C_6$=1.47714$\cdot 10^{7}$ \rcm\AA $^{6}$} &
\multicolumn{2}{l}{$A_{ex}$=3.81419$\cdot 10^{4}$ \rcm\AA $^{-\gamma}$} \\
\multicolumn{2}{l}{$C_8$=4.33209$\cdot 10^{8}$ \rcm\AA $^{8}$} &
\multicolumn{2}{l}{$\gamma$=5.0568} \\
\multicolumn{2}{l}{$C_{10}$=1.21271$ \cdot 10^{10}$ \rcm\AA
$^{10}$}&
\multicolumn{2}{l}{$\beta$=2.2006 \AA $^{-1}$} \\
&&&\\
 \multicolumn{2}{l}{$T^{\mathrm a}_{\mathrm e}$=5566.0898 \rcm} &
{$R^{\mathrm a}_{\mathrm e}$=5.2472 \AA}&\\
 \multicolumn{2}{l}{$D^{\mathrm a}_{\mathrm e}$=309(10) \rcm} &
 \multicolumn{2}{l}{$D^{\mathrm a}_{\mathrm 0}$=287(10) \rcm} \\

 \\\hline
\end{tabular*}
\label{ipapota}
\end{table}

In Tables~\ref{ipapotX} and \ref{ipapota} the fitted potential
energy curves of the X$^1\Sigma^+$ and a$^3\Sigma^+$ states are
given. The dispersion coefficients $C_6$ and $C_8$ are taken from
Refs.~\cite{Derevianko_C62001,Porsev:03}. With the present data
sets we are also able to reproduce the experimental data with the
same quality of the fit by fixing these coefficients to the values
from Ref.~\cite{Marinescu_1999}. The reason for choosing the more
recent values is that in this case the fitted $C_{10}$ coefficient
differs from the theoretical prediction by only -17~\%, whereas if
the leading dispersion coefficients are fixed to the values from
Ref.~\cite{Marinescu_1999} the difference reaches +68~\% (the
derived $C_{10}$ amounts in this case to $2.39 \times
10^{10}$ \rcm\AA$^{10}$). The selected set gives good consistency
with the expected accuracy of most recent calculations of dispersion
coefficients.

The potential curve at any point $R<R_{o}$ is defined by the natural
cubic spline function through \emph{all} points listed in
Tables~\ref{ipapotX} and \ref{ipapota}. For $R \ge R_o$ the long
range parameters and expressions (\ref{LRs}) and (\ref{exchange})
should be used. For convenience of the reader, we give in
Tables~\ref{ipapotX} and \ref{ipapota} also $T_e$ (energy of
potential minimum), $R_e$ (equilibrium distance), $D_e$
(dissociation energy) and the dissociation energy $D_0$ with respect
to the lowest rovibrational level ($v''=0, J''=0$). While the model
parameters for the potential are listed with all relevant figures
necessary to reproduce the model with sufficient precision, for the
dissociation energies as physical quantities uncertainties have been
estimated according to the data situation. Especially for the
a$^3\Sigma^+$ state the uncertainty of the dissociation energy is
fairly large because of only few data, especially no data are
available yet for the lowest vibrational levels.

The derived X$^1\Sigma^+$ state potential describes the
experimental transition frequencies involving 2400
energy levels of the ground state with a standard deviation of 0.0057 \rcm\ and a
dimensionless standard deviation of $\bar{\sigma}=0.62$. The high
standard deviation of the fit compared to the estimated error limits of 0.003 to 0.005 \rcm~from the typically applied experimental resolution of 0.03 - 0.05 \rcm\ arises from a relatively large
number of supplementary spectra (giving rise to about 26 \% of the
identified transitions) recorded at lower resolution (0.1 \rcm)
which are also included in the data analysis. The standard
deviation for the experimental data with uncertainties less than
0.005 \rcm\ (about 3600 transitions) amounts to 0.0030 \rcm\ and
$\bar{\sigma}$ for this case is 0.88. The increase of the
dimensionless standard deviation is most likely due to
overestimated error limits of the low resolution lines. The
quality of the triplet state potential is assessed by comparing
the experimental term energies with the calculated eigenvalues.
The standard deviation amounts to 0.0044 \rcm\ and the
dimensionless standard deviation is 0.51.

In addition to the potential energy curves a set of Dunham
coefficients was fitted to the data for the singlet ground state.
These coefficients are given in Table 5 of the supplementary
materials~\cite{EPAPS} and describe the experimental data for all
rotational quantum numbers and a reduced set of $v_{X}''$: $0\le
v_{X}'' \le 45$.

\section{Conclusion}\label{sec:conclusion}

Highly accurate potentials for the singlet and triplet ground state
were derived, from which one can read off the quality of the ab
initio result \cite{Korek_2000}. Because a graphical comparison is
generally too rough we compare instead two quantities, namely the
dissociation energy $D_e$ and the equilibrium internuclear
separation $R_e$. For the X$^1\Sigma^+$ state we find
$D_e$=5875.455\unit{\rcm} and $R_e$=3.6681\unit{\AA} with the
corresponding ab initio results being 5996\unit{\rcm} and
3.615\unit{\AA}, respectively~\cite{Korek_2000}. For the
a$^3\Sigma^+$ state the present work reveals $D_e$=309\unit{\rcm}
and $R_e$=5.2472\unit{\AA}, while the corresponding ab initio
calculations give, respectively, 307\unit{\rcm} and 5.229\unit{\AA}.

The energy difference in the case of the singlet state could
correspond to at least one vibrational level more than the total
number of levels accommodated in the derived potential (55 for J=0),
but for the triplet state the agreement is surprisingly good. Such
precision is rather good and helpful for guiding the spectroscopic
assignment. The amplitude of the exchange interaction which can be
estimated from the difference of the theoretical asymptotic singlet
and triplet potentials comes close (30\%) to the value from the fit.

Cold collisions were studied for Li+Cs pairs through sympathetic
cooling by Mudrich \emph{et al.}~\cite{mudrich2002:prl} and trap
loss measurements by Schl\"oder \textit{et al.}~\cite{Schloeder}.
The latter work gives loss rates for processes where excited states
are involved and thus cannot be related directly to cold collision
calculations which are now possible with the ground state potentials
reported in this paper. The former work derives the cross section
for elastic scattering of unpolarized atomic pairs in the hyperfine
ground states $f_{Li}=1$ and $f_{Cs}=3$ to be $8(4)\times 10^{-12} $
cm$^{2}$ assuming Wigner's threshold behavior for s-wave scattering,
i.e., the cross section is independent of collision energy. In this
collision process the channels \textit{f}=2, 3, and 4 of the total
atomic angular momentum $\vec{f}=\vec{f}_{Li}+\vec{f}_{Cs}$ are
involved. Using the potentials reported in this work the cross
sections of these elastic channels were calculated for an energy
range up to $100 \mu $K. The values for $f=3$ and $f=2$ are at least
an order of magnitude smaller than that of $f=4$, thus only this
channel should be taken into account for the comparison to the
experimentally derived value which has 50 \% uncertainty. The value
of this cross section varies only from $2.5$ to $2.0\times 10^{-12}$
cm$^{2}$ from zero to $100 \mu$K energy. Thus the threshold law is
sufficiently well fulfilled. By weighting the cross section of $f=4$
with the statistical weight 9/(9+7+5) according to all existing
channels we get as cross section of the unpolarized collision
$1.1\times 10^{-12}$ cm$^{2}$, which is about a factor 8 smaller
than the one derived from the experiment, but within two times the
given error. The potential for the singlet ground state is well
determined by a large body of data (2397 levels), but the triplet
ground state was only determined by 89 levels. Thus we believe that
the difference is not directing to a discrepancy between both
results but ask for more spectroscopic data or more precise cross
section measurements or a direct observation of Feshbach resonances
which can be incorporated in the fit of potential functions to
obtain a full description of the spectroscopy and the cold
collisions of Li + Cs atom pairs. Calculations of scattering length
for the singlet and triplet states show that for the singlet ground
state it is already well determined by the present study and will be
50(20) a$_0$ (atomic unit a$_{0}=0.529\times 10^{-10}$ m), but in
the case of the triplet state large values with different signs were
obtained during the evaluation with potentials represented with
spline coefficients or with piecewise analytic functions as used in
some of our other work (see e.g.~\cite{Docenko_NaCs2006}). Thus,
such values are not yet reliable and calculations of Feshbach
resonances would be of no value. To make such calculations reliable
more spectroscopic data for the low vibrational levels of the
a$^3\Sigma^+$ state and of near asymptotic levels of both ground
states is needed. In addition to collecting data by exciting to high
lying levels of the B$^1 \Pi$ state we searched for, but did not yet
find, other excitation channels which will lead to combined
fluorescence to both ground states, especially to asymptotic levels.
For improving predictions of appropriate excitations we started new
experiments to get precise data of various excited states. The
analysis of the states B$^1 \Pi$ and D$^1 \Pi$ is almost complete
and will be published in a forthcoming paper \cite{pashovBstate}.

\begin{acknowledgments} This work was supported by the Deutsche
Forschungsgemeinschaft in the frame of the Sonderforschungsbereich
407 and by the European Commission in the frame of the Cold
Molecule Research Training Network under contract
HPRN-CT-2002-00290. A.P. acknowledges partial support from the
Bulgarian National Science Fund grant MUF 1560/05.
\end{acknowledgments}


\end{document}